# Anisotropic Impurity-States, Quasiparticle Scattering and Nematic Transport in Underdoped Ca(Fe$_{1-x}$Co$_x$)$_2$As$_2$

M. P. Allan[1,2,3], T.-M. Chuang[2,3,4,8], F. Massee[2,3,5], Yang Xie,[2] Ni Ni,[6,7] S. L. Bud'ko,[6,7] G. S. Boebinger,[8] Q. Wang,[9] D. S. Dessau,[9] P. C. Canfield,[6,7] M. S. Golden[5] and J. C. Davis[2,3,10,11]

1  Department of Physics, ETH Zurich, CH-8093 Zurich, Switzerland
2  LASSP, Department of Physics, Cornell University, Ithaca, NY 14853, USA
3  CMPMS Department, Brookhaven National Laboratory, Upton, NY 11973, USA
4 Institute of Physics, Academia Sinica, Nankang, Taipei 11529, Taiwan
5  Van der Waals-Zeeman Institute, University of Amsterdam, 1098 XH Amsterdam, The Netherlands
6  Ames Laboratory, U.S. Department of Energy, Iowa State University, Ames, IA 50011, USA
7 Department of Physics and Astronomy, Iowa State University, Ames, IA 50011, USA
8 NHMFL, Department of Physics, Florida State University, Tallahassee, FL 32310, USA
9 Department of Physics, University of Colorado, Boulder, CO 80309, USA
10 SUPA, School of Physics and Astronomy, University of St Andrews, St Andrews, Fife KY16 9SS, UK
11 Kavli Institute at Cornell for Nanoscale Science, Cornell University, Ithaca, NY 14850, USA

**ABSTRACT**  Iron-based high temperature superconductivity develops when the 'parent' antiferromagnetic/orthorhombic phase is suppressed, typically by introduction of dopant atoms[1,2]. But their impact on atomic-scale electronic structure, while in theory quite complex[3-14], is unknown experimentally. What is known is that a strong transport anisotropy[15-26] with its resistivity maximum along the crystal *b*-axis[15-26], develops with increasing concentration of dopant atoms[15,21-26]; this 'nematicity' vanishes when the 'parent' phase disappears near the maximum superconducting T$_c$. The interplay between the electronic structure surrounding each dopant atom, quasiparticle scattering therefrom, and the transport nematicity has therefore become a pivotal focus[8,9,13,23,24] of research into these materials. Here, by directly visualizing the atomic-scale electronic structure, we show that substituting Co for Fe atoms in underdoped Ca(Fe$_{1-x}$Co$_x$)$_2$As$_2$ generates a dense population of identical anisotropic impurity states. Each is ~8 Fe-Fe unit cells in length, and all are distributed randomly but aligned with the antiferromagnetic *a*-axis. By imaging their surrounding interference patterns, we further demonstrate that these impurity states scatter quasiparticles in a highly anisotropic manner, with the maximum scattering rate concentrated along the *b*-axis. These data provide direct support for the recent proposals[8,9,13,23,24] that it is primarily anisotropic scattering by dopant-induced impurity states that generates the transport nematicity; they also yield simple explanations for the enhancement of the nematicity proportional to the dopant density[15,21-26] and for the occurrence of the highest resistivity along the *b*-axis[15-26].

Commensurate antiferromagnetism with orthorhombic crystal symmetry exists in the phase 'parent' to the superconductivity in most underdoped FeAs materials[1,2]. The $a$-axis unit cell is fractionally longer than that of the $b$-axis, with the Fe spins aligning in an anti-parallel fashion along the $a$-axis but parallel along the $b$-axis. Doping is usually achieved by substitution of $x$ transition-metal atoms per Fe. When $x \sim 4 \pm 1\%$, the superconductivity appears, reaching its maximum $T_c$ usually near $x_c \sim 8 \pm 1\%$ (Fig. 1a) where the antiferromagnetic/orthorhombic phase disappears. A broken discrete rotational symmetry of both the crystal structure and the electronic characteristics, usually referred to as 'nematicity', precedes or coincides with the appearance of antiferromagnetism (inset Fig. 1a from Ref. 15). Various techniques including spectroscopic imaging STM[27-30], DC electrical transport[15-17,20-23,25,26,34] (DCT), frequency-dependent electrical conductivity[18,19,24,34] (ACT), torque magnetometry[31], and angle resolved photoemission[32-34] (ARPES) now support this picture of intense, but unexplained, electronic nematicity in underdoped FeAs materials.

Before bulk detwinning of FeAs crystals was available, the only way to search for electronic nematicity was by atomic-scale electronic structure imaging. In $Ca(Fe_{1-x}Co_x)_2As_2$ (Refs 35-37), this approach allowed discovery and visualization of the electronic nematicity[27]. But, unexpectedly, it coexists with intense unidirectional nanoscale electronic disorder aligned with the $a$-axis. When detwinning techniques were developed, the DCT studies in $BaFe_2As_2$ (Ba-122) revealed strikingly anisotropic electronic transport[15,16]. This transport anisotropy is actually very weak at $x=0$, grows towards a maximum near $x \sim 4\%$, and disappears beyond $x=x_c$ (Fig. 1a), a situation now widely observed in these materials[15,21-23,25,26,34]. The higher resistivity (quasi-insulating) axis is always the ferromagnetic $b$-axis[15-17,20-23,25,26], as also detected in high frequency transport studies[18,19,24]. However, these anisotropic transport characteristics cannot be due merely to crystal orthorhombicity. Firstly, they are minimal at zero doping where orthorhombicity is a maximum. More compelling are the discoveries that residual resistivity[23,24], as well as the resistivity anisotropy, increase ~linearly with Co dopant density[15,21-26], and, moreover, that atomic substitutions outside the FeAs plane generate very weak transport anisotropy[22]. Therefore, an additional hypothesis has recently been advocated: that the anomalous doping dependence of nematic transport in underdoped FeAs materials occurs primarily because of anisotropic scattering generated by the dopant atoms[8,9,13,23,24]. Thus, the structure of impurity-states generated by dopant atoms[3,4,5-7,9,13], how this influences the scattering of quasiparticles[6,8,9,13], and the relationship between these atomic-scale phenomena and the transport anisotropy[8,9,13,23,24] have become a key focus of research into the phase diagram of these compounds.

Testing the above hypothesis requires simultaneous atomic-scale visualization of the impurity states and the resulting quasiparticle scattering. This we achieve by using spectroscopic imaging STM to study Ca(Fe$_{1-x}$Co$_x$)$_2$As$_2$ samples with 0≤$x$~6%. This material shows the expected suppression of antiferromagnetism and the emergence of superconductivity (often without coexistence), but precise phase diagram details depend on annealing. These are cleaved in cryogenic ultrahigh vacuum and then inserted into the STM head; all data reported here were acquired at 4.2 K. The surfaces are atomically flat and exhibit a 1×2 surface reconstruction at ~45° to both the *a*- and *b*-axes which we have previously demonstrated to have no impact on accurate visualization of FeAs plane electronic structure[27]. We image the differential tunneling conductance $dI/dV(\boldsymbol{r}, E = eV) \equiv g(\boldsymbol{r}, E)$ with atomic resolution and register, and as a function of both location $\boldsymbol{r}$ and electron energy $E$. The Fourier transform of $g(\boldsymbol{r}, E)$, $g(\boldsymbol{q}, E)$, is then used to determine the characteristic wavevectors of dispersive modulations due to quasiparticle scattering interference (QPI). As shown in Fig. 1b, these patterns exhibit[27] a strongly unidirectional hole-like dispersion along the *b*-axis only, with two apparent "satellites" of the basic unidirectional dispersion characteristics shifted by δ$\boldsymbol{q}$≅±2π/8a$_0$ along the *a*-axis. Importantly, without the Co-dopant atoms at $x$=0, these QPI effects are absent (Ref. 27 and Supplementary Information (SI) Section I). At finite doping, the observed quasiparticle interference pattern (Fig. 1b) is highly inconsistent with predictions using the relevant Fermi surface of the same material[32] shown in Fig. 1c (see Fig. 3a,b, and SI Section II). Finally, if one interpreted these observations as the preferential propagation of delocalized electrons along the *b*-axis, this would be highly inconsistent with the DCT and ACT studies[15-26] reporting that the low-resistance transport always occurs along the *a*-axis. An explanation for all these apparently contradictory phenomena is required.

These QPI effects coexist with intense but poorly understood static electronic disorder at the nanoscale[27]. Figure 2a shows a typical topographic image of the cleaved surface of underdoped Ca-122 while Fig. 2b is the simultaneously acquired current image of non-dispersive states achieved by integrating the electronic structure images over energy $I(\mathrm{r}, -37meV) = \int_0^{E \approx -37meV} g(\mathrm{r}, \omega) d\omega$ to remove dispersive effects. This shows vividly the energetically quasi-static electronic disorder strongly oriented along the *a*-axis. To explore the nanoscale constituents of this disorder, we carry out an autocorrelation analysis of $I(\boldsymbol{r}, V)$ as shown in Fig. 2c. Although this reveals no periodic modulations in the static electronic disorder, there is a sharply defined structure at the center of the autocorrelation. The inset to Figure 2c showing its full details reveals that, beyond the required central peak, there are two peaks along the *a*-axis which are separated from the central point by a distance of $|\mathbf{d}|$ = 22Å ≈ 8$a_0$ (SI Section II). The absence of other features in the autocorrelation demonstrates that $|\mathbf{d}|$ is the only persistence length. Figure 2d shows the Fourier transform of Fig. 2b from which we conclude that there is no periodicity whatsoever (thus excluding periodic 'stripes' as the explanation).

We next consider what type of nanoscale phenomenon could generate all the effects in Fig. 2b-d. We analyze a model hypothesizing identical, co-oriented but otherwise randomly located, anisotropic impurity states. The inset to Fig. 2e shows a model impurity state in real-space in the shape of an "electronic dimer". For reasons clarified below, it consists of two Gaussians $G_\sigma(\mathbf{r}) = g_0 \left( e^{-|\mathbf{r}|^2/\sigma^2} \right)$ separated from each other by a distance $|\mathbf{d}|$ = 22Å ≈ 8$a_0$ where $\mathbf{d}$ is always parallel to the *a*-axis, so that the complete impurity state is represented by $D(\mathbf{r}) = G_\sigma(\mathbf{r} + \mathbf{d}/2) + G_\sigma(\mathbf{r} - \mathbf{d}/2)$. We then consider a randomly distributed ensemble of such co-oriented anisotropic impurity states, as would be observable in the energy-integrated tunnel-current map $I(\mathbf{r})$. The left-hand panel of Fig. 2e shows four such impurity states at random locations exemplifying the basic structure of our model. The right-hand panel shows the effect of ~47 such states distributed at random. Figure 2f shows results from precisely the same model but now with 1000 *a*-axis oriented anisotropic impurity states - the approximate number of Co dopant atoms within this field of view at $x$~3.0% (SI section IV). A comparison with Fig. 2b reveals the remarkable consistency between the predictions from this model and our experimental observations. Figure 2g is the autocorrelation of Fig. 2e and exhibits the identical central feature. The comparison between the central features in inset Fig. 2c (experiment) and inset Fig. 2g (model) is very good. Finally Fig. 2h shows the Fourier transform of the simulation in Fig. 2f; it lacks any peaks that would indicate a periodic structure. Once again, the correspondence between the experimental result in Fig. 2d and the model simulation in Fig. 2h is excellent. In summary: the correspondences between Fig.'s 2b and 2f, 2c and 2g, and 2d and 2h respectively, provides compelling evidence that the static electronic disorder in underdoped Ca-122 indeed consists of a randomly distributed ensemble of *a*-axis oriented electronic anisotropic impurity states, all of size $|\mathbf{d}|$ = 22Å ≈ 8$a_0$.

Another key observation is shown in the inset to Fig. 2b (SI Section V,VI). Here the electronic structure surrounding every observed Co dopant-atom location[3,27] is demonstrated. This is identified by averaging $I(\mathbf{r}, V)$ over many small fields of view, each being centered on a Co substitution site, to yield $\langle I(\mathbf{r}, V) \rangle_{Co}$. We see directly that two peaks exist in $\langle I(r,V)\rangle_{Co}$, each separated from the dopant-atom site (red +) by $\pm d/2$. These data provide direct atomic-scale evidence that Co dopant-atoms play a central role in either the establishment or pinning of the observed anisotropic impurity states.

The existence of these anisotropic impurity states can be expected to influence various aspects of the physics of underdoped FeAs materials. Perhaps most importantly, it has been proposed that such impurity states scatter quasiparticles anisotropically and thus contribute significantly to the nematic transport[8,9,13,23,24]. Quite generally, electronic transport can be described by the Boltzmann transport equation, which allows for two symmetry breaking terms: (i) the Fermi velocities, $v_F(\mathbf{k})$, and (ii) the scattering probabilities $V(\mathbf{k},\mathbf{k'})$. Some proposals have concentrated on the Fermi velocity anisotropy, and assumed that a rigid band shift leads to a change thereof upon doping[21]. Here, however, we explore the complementary proposals[9,13,23,24] that it is anisotropic impurity scattering which contributes significantly the nematic transport. Quasiparticle interference imaging can evaluate such proposals because it directly measures the scattering of quasiparticles. For a point scatterer in the Born approximation, the power-spectral-density in differential conductance modulations due to QPI, $P(\mathbf{q},\omega) \propto g(\mathbf{q},\omega)$, can be determined from[39]

$$P(\mathbf{q},\omega) = \frac{1}{N}\left|\frac{1}{\pi}\mathrm{Im}\Lambda(\mathbf{q},\omega)\right|^2 |\delta\varepsilon(\mathbf{q})|^2 \qquad (1)$$

Here the term $\left|\frac{1}{\pi}\mathrm{Im}\Lambda(\mathbf{q},\omega)\right|^2$, with $\Lambda$ being a specific convolution of Greens functions $\Lambda(\mathbf{q},\omega) \equiv \int d^2r\ e^{i\mathbf{q}\cdot\mathbf{r}} G(\mathbf{r},\omega)G(\mathbf{r},-\omega)$, represents the bare QPI signature expected from a point scatterer within the native band structure. The $|\delta\varepsilon(\mathbf{q})|^2$ term is the square of the structure factor (Fourier transform of the $\mathbf{r}$-space scattering potential) of any non-point-like scattering center. Eqn. 1 shows how such an anisotropic scatterer would influence the QPI data.

If the spatial extent of each scatterer is modeled by a delta function in $\mathbf{r}$-space and the measured $A(\mathbf{k},\omega)$ of Ca-122 is used to model Fermi surface and band dispersions (SI Section II), this predicts the $P(\mathbf{q},\omega=0)$ as shown in Fig. 3a. In Fig. 3b, we show the measured $g(\mathbf{q},E=0)$ that are obviously strongly inconsistent with Fig. 3a. One might therefore be tempted to deduce that the reported QPI data are artifacts unrelated to the fundamental FeAs band structure. However, if the scattering centers exhibit the structure factor of the model "electronic dimer" impurity state (Fig. 2e) :

$$|\delta\varepsilon(\boldsymbol{q})|^2 = FT\{G_\sigma(\boldsymbol{r}-\boldsymbol{d}/2)+G_\sigma(\boldsymbol{r}+\boldsymbol{d}/2)\} \propto 2|\cos(\boldsymbol{d}.\boldsymbol{q}/2)|G_{1/\sigma}(\boldsymbol{q}) \quad (2)$$

we find a very different conclusion. Figure 3c shows the predicted QPI patterns for underdoped Ca-122 again using the native $A(\boldsymbol{k},\omega)$ as measured by ARPES to predict $\Lambda(\boldsymbol{q},\omega)$. Here we find three parallel lines separated from each other by $\delta\boldsymbol{q}\cong\pm 2\pi/8a_0$ and all dispersing identically on the same trajectory of Fig. 1b (or red line in Fig. 3d), so that the equivalent agreement between Fig.'s 3b and 3c holds for all energies. Thus we show that, for Ca-122, the $A(\boldsymbol{k},\omega)$ from ARPES and the observed QPI are actually quite consistent with each other if dopant-induced $\sim 8a_0$ anisotropic impurity states act as the predominant scatterers.

Conversely, QPI measurements can be used to measure directly the actual scattering characteristics of impurity states in Ca-122. Figure 4b shows the angular dependence of the quasiparticle scattering rate determined using the square of the structure factor of the observed anisotropic impurity states (Fig. 4a). It is clearly highly anisotropic and its maximum $\boldsymbol{q}$-integrated intensity is found to occur along the $b$-axis. While such applications of the Boltzmann transport equation can be complex and do depend on the Fermi surface, our analysis nevertheless provides a plausible explanation for how the scattering rates of the anisotropic impurity states (Fig. 2,3) affect the transport nematicity. Further, it provides a simple explanation for why the expected easy transport $b$-axis actually exhibits the highest resistivity.

Overall, our data demonstrate that insertion of Co dopant atoms on the Fe sites in Ca-122 generates a dense ensemble of $\sim 8a_0$ long highly anisotropic impurity states, which, because of their highly oriented structure, may be locally magnetic in origin[6,7]. Notwithstanding their cause, the measured characteristics of these impurity states then provide a simple yet comprehensive explanation for all the complexities in the observed QPI phenomenology[27]. And, perhaps most importantly, our studies show directly that quasiparticle scattering by these impurity states is actually highly anisotropic. Therefore, we demonstrate that anisotropic scattering by impurity states can provide a simple, comprehensive explanation (Fig. 4b) for the complex features of the transport nematicity in iron-arsenides, including its microscopic source, why its high resistivity is focused along what should be the easy transport $b$-axis, and why it only becomes robust with increased dopant atom concentration.

**Figure Captions**

Figure 1    Anisotropic transport and quasiparticle scattering interference in underdoped iron-based superconductors
a. (*A*) Schematic phase diagram of $A(Fe_{1-x}Co_x)_2As_2$ (A=Ca, Sr and Ba) as a function of doping concentration x including the structural, magnetic, and superconducting transition temperatures[1,35-38]. The dashed line indicates nematic fluctuations. Inset: The evolution of the in-plane resistivity anisotropy as a function of temperature and doping, expressed in terms of the resistivity ratio $\rho_b/\rho_a$ (reproduced from Ref. 15).
b. The unidirectional dispersion of the six QPI peaks occurs along the *b*-axis only[27] with two "satellites" of the central unidirectional dispersion shifted by $\Delta q \approx 1/8a_0$.
c. Photocurrent intensity at the Fermi level (E=0) of Ca-122 measured by ARPES (from Ref. 32).
Panels b,c are oriented so that the crystal *b*-axis points vertically.

Figure 2    Visualizing the anisotropic impurity state structure in Ca-122
a. A 48x48nm$^2$ topographic image of the Ca-122 surface, taken simultaneously with the measurements depicted in (b-d). The horizontal lines stem from a surface reconstruction and do not influence our ability to visualize the correct FeAs electronic structure; the orientations of crystal axes are identified.
b. The simultaneously recorded non-dispersive component in electronic structure as determined the current map I(*r*,E) at E=-37meV. The inset shows the characteristic non-dispersive electronic structure environment of a typical Co dopant atom (see SI Section V,VI): it is observed directly to be a "dimer" shaped electronic impurity state.
c. The autocorrelation of I(*r*,-37meV) shows three peaks separated by the characteristic length scale of 22Å. Note that apart from the triple-peak, the AC{*I*(*r*,-37meV)} signal is low, proving that 22Å is the *only* persistent length scale in the image.
d. Fourier transform I(*q*,-37meV) of the image shown in b. No sharp peak indicating a periodic structure are seen anywhere in reciprocal space.
e. Inset, the proposed impurity state with the Fe-lattice for comparison. The lower half show simulations with n=5 and n=47.
f. A simulation with n=1000 anisotropic impurity states in the same 48x48nm$^2$ FOV (SI Section IV); their centers are randomly distributed but they are all aligned with the a-axis. This 'glassy' pattern of overlapping anisotropic impurity states looks similar to the data shown in b-d.  The autocorrelation and Fourier transform (g,h) of the image in f support the validity of  our deduction that the static electronic disorder consists of *a*-axis oriented electronic anisotropic impurity states only.

Figure 3    Anisotropic quasiparticle scattering interference from anisotropic impurity states
(See SI Section VII for a comparison of **q**- and **k**-space.)
a.  The QPI pattern one would expect from the ARPES band structure (Fig. 1e and SI Section II) with the quasiparticles scattering from point-like scatterers.
b.  The QPI pattern as measured in Ca-122 in the same energy range. While the dispersion is correct, there are clear vertical lines of suppressed intensity in the data that is not visible in the simulations.
c.  Using the structure factor of a "dimer" (Fig. 2) impurity state, the calculations yield QPI predictions consistent with the data in b.
d.  The dispersion of the measured QPI (red line) is in agreement with the dispersion determined by ARPES (photocurrent intensity in gray scale).

Figure 4    Anisotropic transport due to scattering from anisotropic impurity states
a. False color plot of the probability for a state **k** to scatter into a state **k'**, which in the Born approximation using plane waves is proportional to the Fourier transform of the scattering potential $W(\mathbf{k},\mathbf{k'}) = W(\mathbf{q}) \sim |FT\{V(\mathbf{r})\}|^2$. We take the scattering potential to have the same spatial form as the observed anisotropic impurity states (Fig. 2), which then leads to the depicted $C_2$-symmetric scattering probability $W(\mathbf{q}=\mathbf{k}-\mathbf{k'})$.
b. A schematic phase diagram with increasingly nematic transport is indicated by blue, yellow and red colors, where the latter marks strongest transport anisotropy. To emphasize the directional dependence of the scattering probability from our QPI data, we plot in the inset the cumulative scattering probability in a given direction $\theta = \angle(k,k')$. This is obtained by integrating over the radial coordinate: $W(\theta) = \int d|\mathbf{q}| W(\theta,|\mathbf{q}|)$ in a. Note the strong scattering maxima along the *b*-axis.

**Acknowledgments:** We acknowledge and thank F. Baumberger, P. Dai, P.J. Hirschfeld, J. E. Hoffman, A. Kaminski, E.-A. Kim, D.-H. Lee, J. Orenstein, G. Sawatzky, D. J. Scalapino, J. Schmalian, and S. Uchida for helpful discussions and communications. We are especially grateful to N.D. Loh for insightful suggestions and to A. W. Rost for proposing the analysis and presentation in Fig. 4b. The reported studies are supported by the Center for Emergent Superconductivity, a DOE Energy Frontier Research Center headquartered at Brookhaven National Laboratory. Work at the Ames Laboratory are supported by the DOE, Basic Energy Sciences under Contract no. DE-AC02- 07CH11358. Further support comes from NSF/DMR-0654118 through the National High Magnetic Field (T.-M.C), the Cornell Center for Materials Research under NSF/DMR-0520404 (Y.X.); the U.K. Engineering and Physical Sciences Research Council and the Scottish Funding Council under the PhD plus program (M.P.A.); and the Foundation for Fundamental Research on Matter (FOM) of the Netherlands Organization for Scientific Research (F.M. and M.S.G.). The authors wish to dedicate this study to the memory of Prof. Zlatko Tešanović, some of whose recent research was focused on issues addressed herein.

# Figure 1

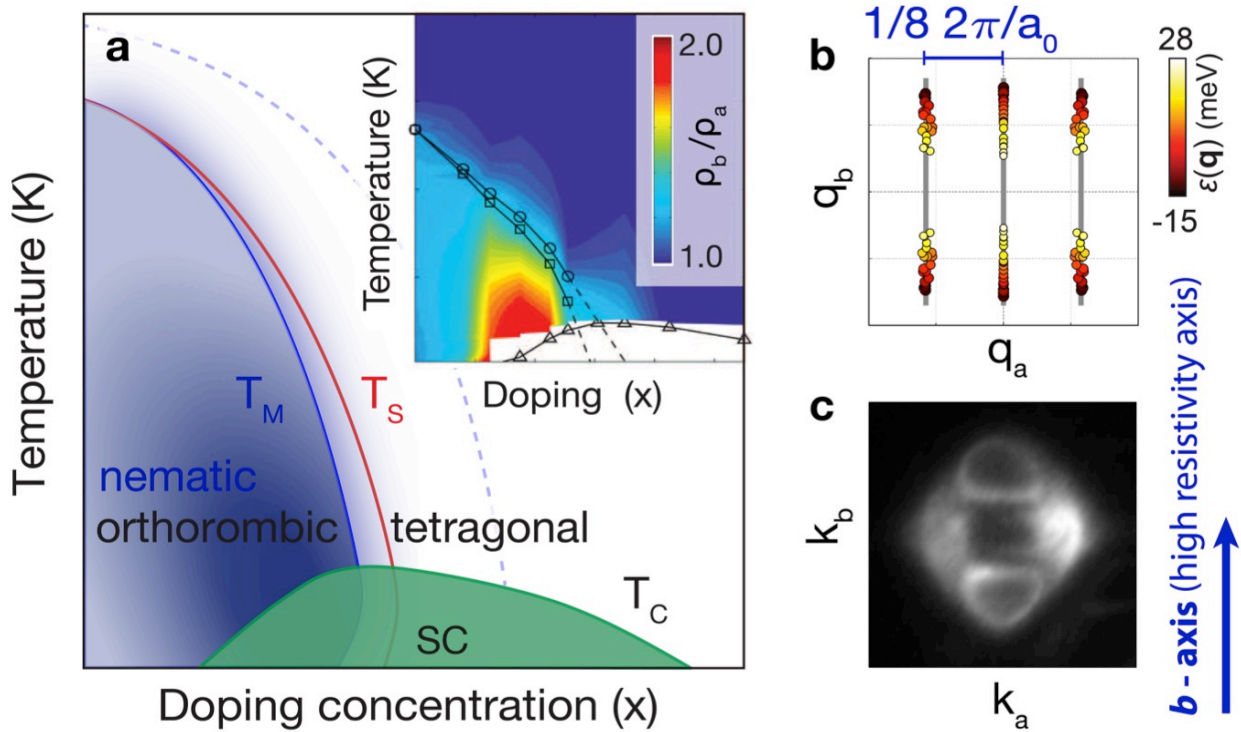

# Figure 2

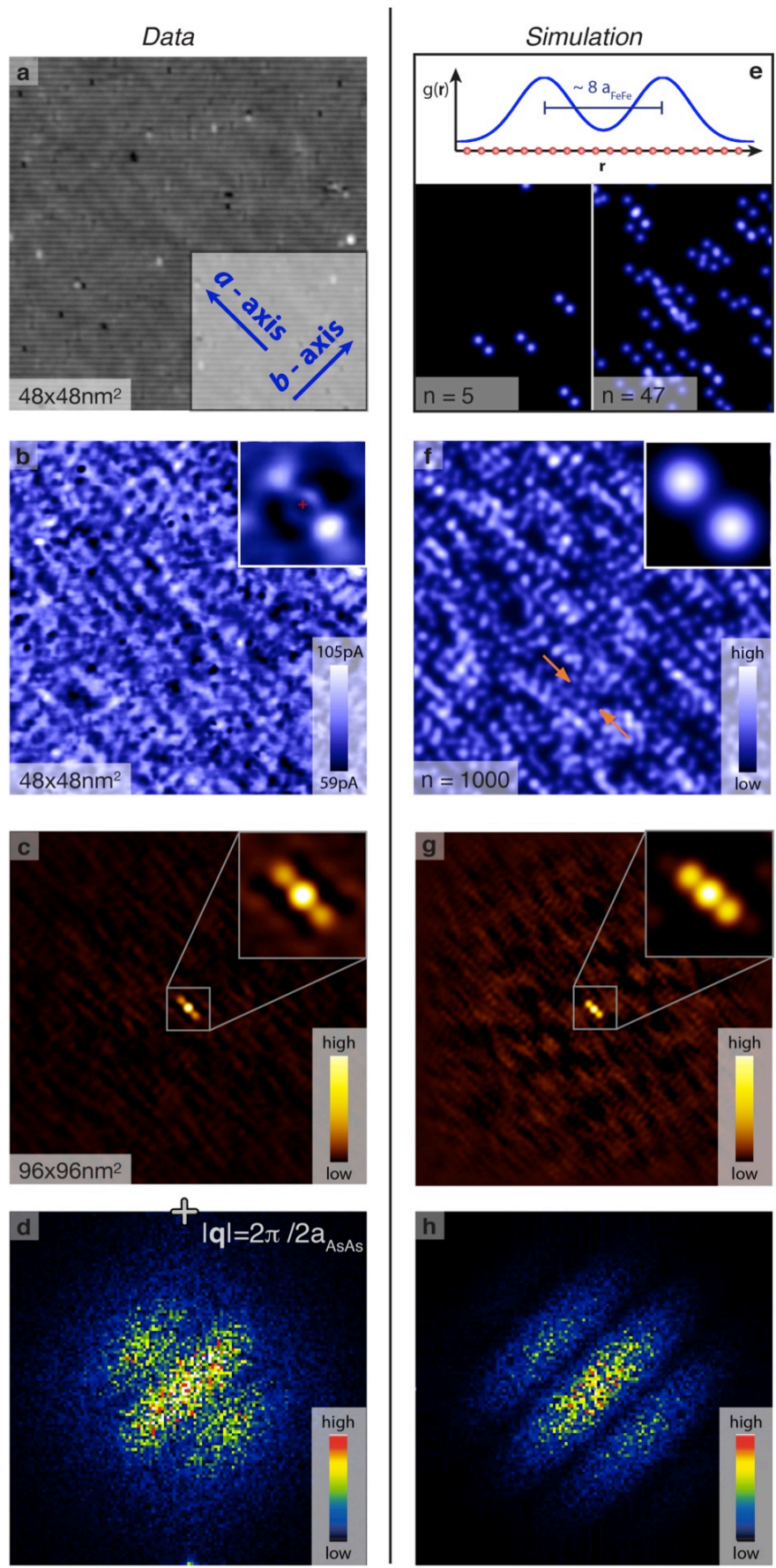

*Figure 3*

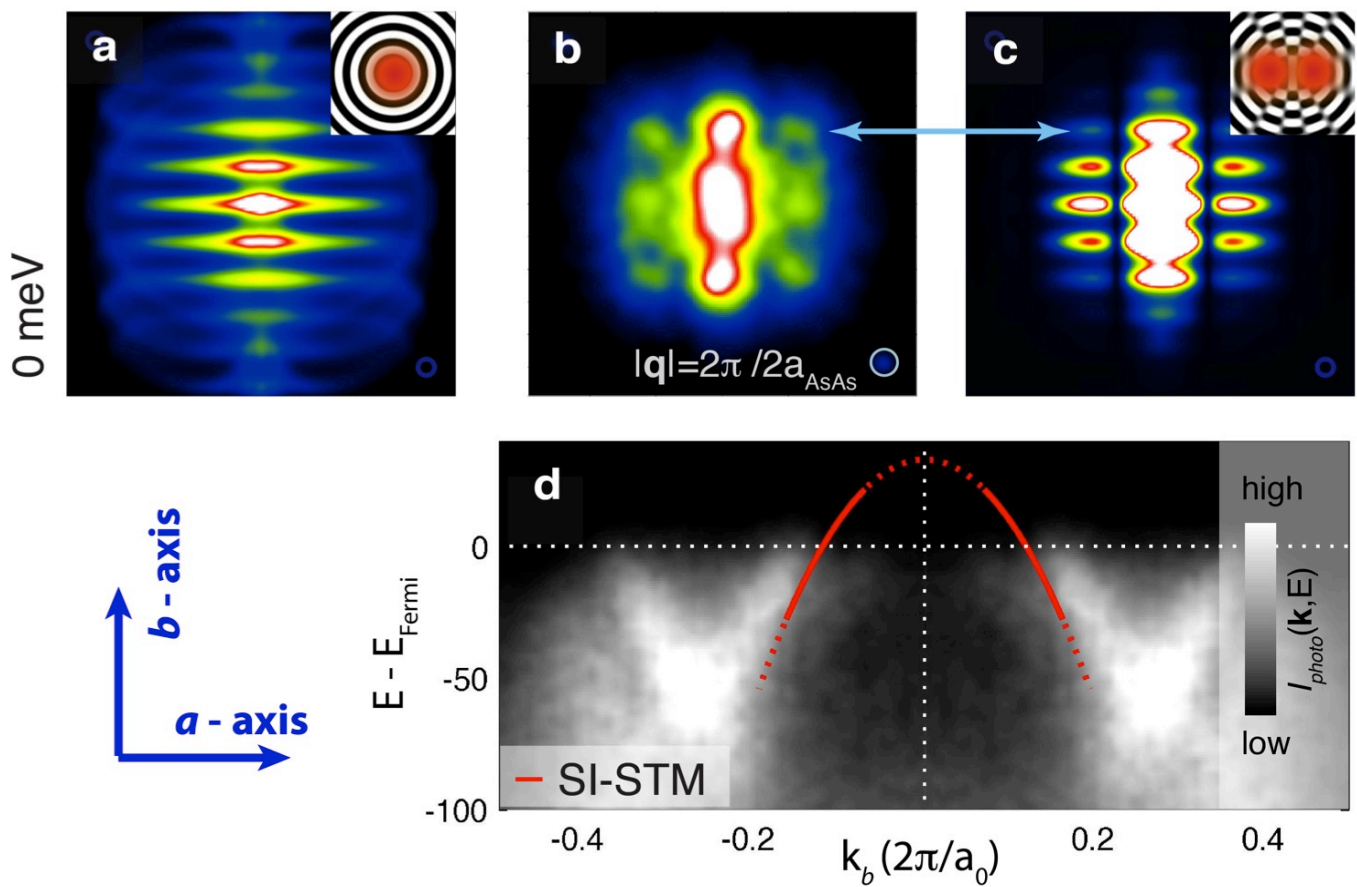

*Figure 4*

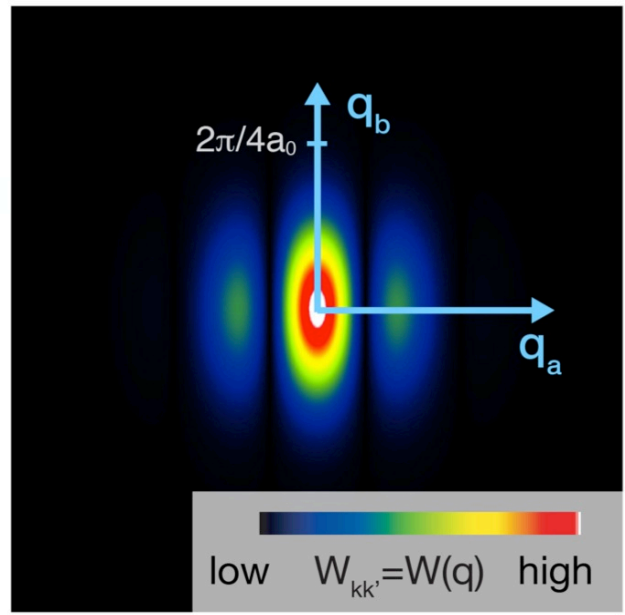

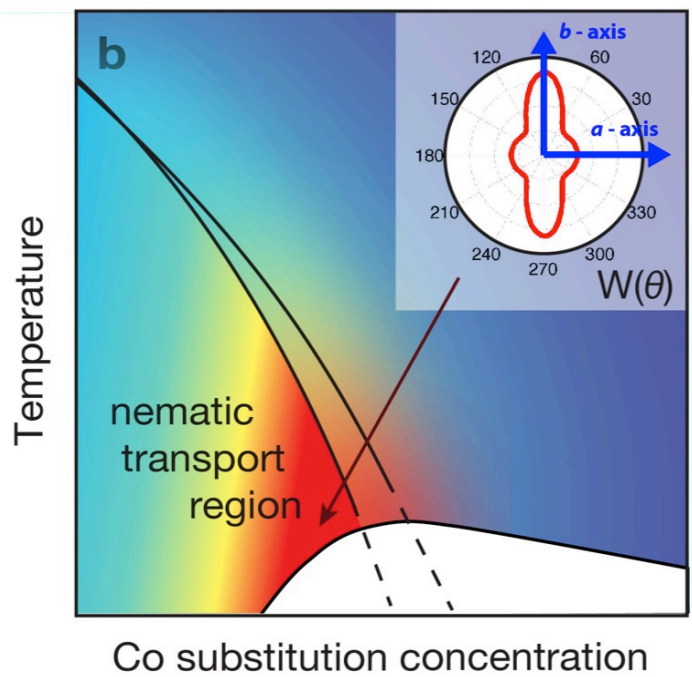

**Supporting Information**

**Anisotropic Impurity-States, Quasiparticle Scattering and Nematic Transport in Underdoped Ca(Fe$_{1-x}$Co$_x$)$_2$As$_2$** by M.P. Allan et al.

I. Doping Dependence of the Electronic Structure in Ca(Fe$_{1-x}$Co$_x$)$_2$As$_2$
II. Modeling QPI signal based on ARPES data
III. Analysis of the autocorrelation AC{g(E= -37, *r*)}
IV. Simulations of random, a-axis oriented anisotropic impurity states
V. Identification of cobalt dopant atoms
VI. Cobalt sites and nanoscale anisotropic electronic impurity states
VII. Comparison of different *q*-vectors observed in QPI pattern

References

## I. Doping Dependence of the Electronic Structure in Ca(Fe$_{1-x}$Co$_x$)$_2$As$_2$

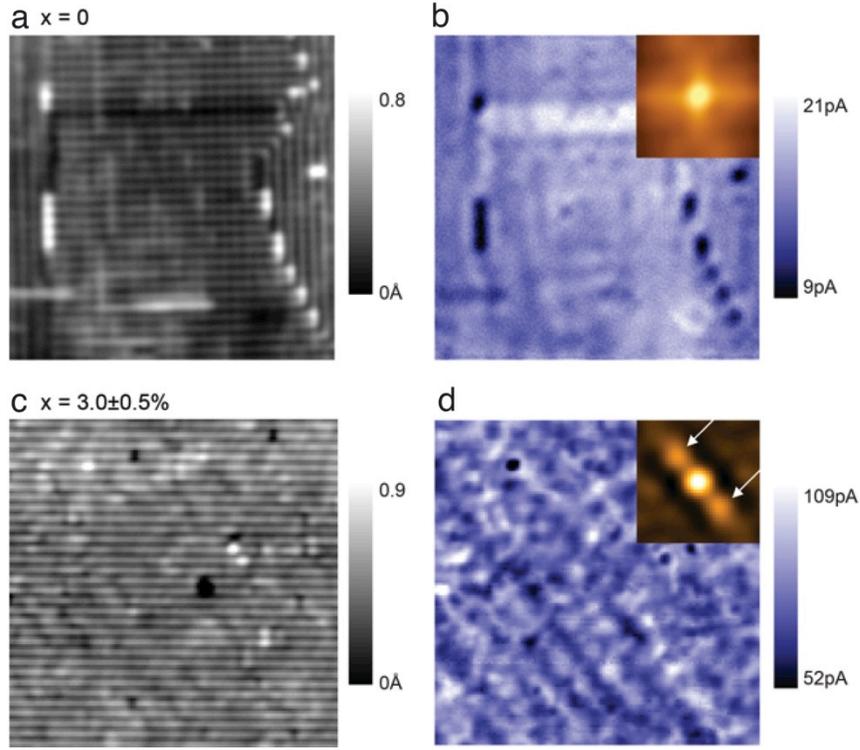

**Fig. S1** Constant-current topographic images of Ca(Fe$_{1-x}$Co$_x$)$_2$As$_2$ taken in a 25nm-square field of view for samples with doping levels of x=0 and x=3.0±0.5% are shown in (a) and (c), respectively. (b) and (d) show current maps, $I(r, E = -36 meV)$ taken simultaneously in the same area for x=0 and x=3.0±0.5%, respectively. The inset in (b) and (d) are the autocorrelation of the current maps.

Topographs of samples with both doping levels (x=0 and x=3.0±0.5%) show the same 1×2 surface reconstruction (its orientation has no impact on the electronic structure shown in the current maps[1]). On the other hand, inserting cobalt atoms produces the profound unidirectional nanoscale electronic structures described in the main text (Fig. S2d). The autocorrelation analysis (inset of Fig. S1d) shows the characteristic length of anisotropic impurity states, 8$a_0$≈22Å, as indicated by the white arrows. Such a feature is absent in the autocorrelation analysis of the undoped sample, where the electronic structure is fairly homogeneous. The only dark contrast shown in current maps coincides with the bright regions in the topographs, which might be Ca clusters on the surface after the violent cryogenic cleave.

## II. Modeling QPI signal based on ARPES measurements

In order to obtain an estimate for the QPI pattern one would expect in $CaFe_2As_2$, we modeled the low energy band structure based on ARPES results. The model-bands $e_\mathbf{k}$ are based on Ref. 2; they are constructed to have the dispersion and Fermi wave vectors measured in the photoemission experiments (Fig. S6). From the model dispersion $e_\mathbf{k}$ we calculated the spectral function $A_{model}(\mathbf{k}, \omega=0meV)$ using a constant imaginary self energy of $\Sigma_{im}=6meV$, i.e $A(k,\omega) \propto \Sigma_{im}/\left[(\omega-\varepsilon_k)^2 + \Sigma_{im}^2\right]$ [3]. Following Ref. 4 and 5, the autocorrelation is then calculated from the spectral function shown in **c**. This is what we can expect, in a simple picture, from a QPI measurement that involves point scatterers. The autocorrelation is proportional to the $\left|\frac{1}{\pi}Im\Lambda(q,\omega)\right|$ term in the main text.

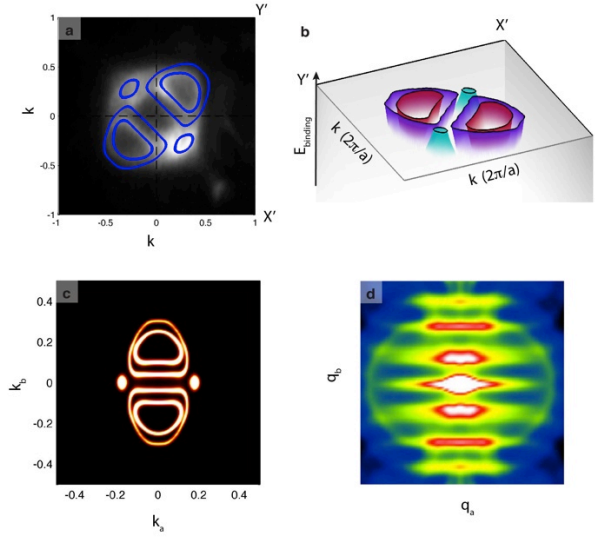

**Fig. S6** Constructing a model band structure based on ARPES results. (a) The model Fermi surface (blue) is overlaid on the photocurrent intensity, c.f. Fig. 4e in Ref. 2. (b) A three-dimensional rendering of the bands below the Fermi level, the black line indicates the Fermi surface. (c) The model spectral function $A_{model}(\mathbf{k}, \omega=0meV)$ calculated using a constant imaginary self energy of $\Sigma_{im}=6meV$. (d) The autocorrelation of the spectral function, which is approximately what one would expect from a QPI measurement from point scatterers.

## III. Analysis of the autocorrelation AC{g(E= -37, r)}

To obtain the parameters |**d**|, σ for our model presented in Fig. 2 *E-G*, we analyzed cuts through peaks in the autocorrelation as depicted in Fig. 2*C*. Fig. S2 shows such a cut. A fit with three Gaussians yields a distance |**d**|≈21Å, the numbers ranges from about 19 to 24 Å over an energy scale of ±30meV. For more information, see Ref. 6.

.

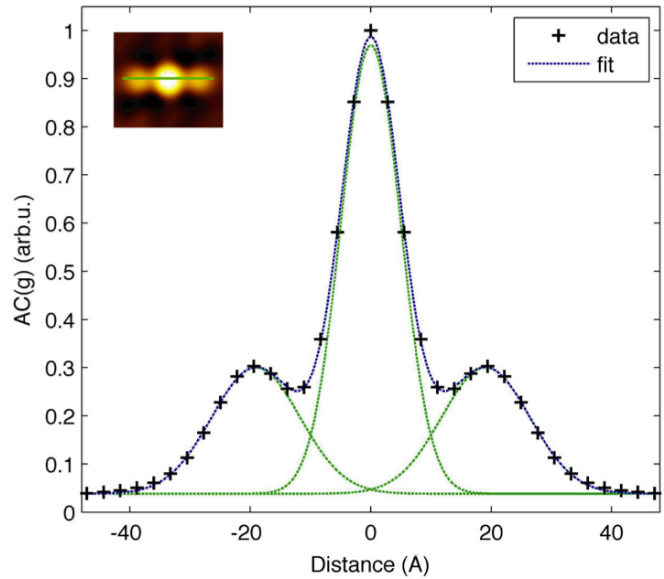

**Fig. S2** A cut through the autocorrelation AC{*g*(E=-37,**r**)} showing the central peak as well as the two side peaks. The location of the cut is marked by a green line in the inset. The shape can be fitted to three Gaussians; the distance between them gives the intra- anisotropic impurity states -distance. At this energy, the fit yields |**d**|≈21Å.

### IV. Simulations of randomly scattered, a-axis oriented anisotropic impurity states

Fig. S3 shows simulations similar to the one depicted in Fig. 2*F*, but with different numbers of randomly scattered, *a*-axis oriented anisotropic impurity states.

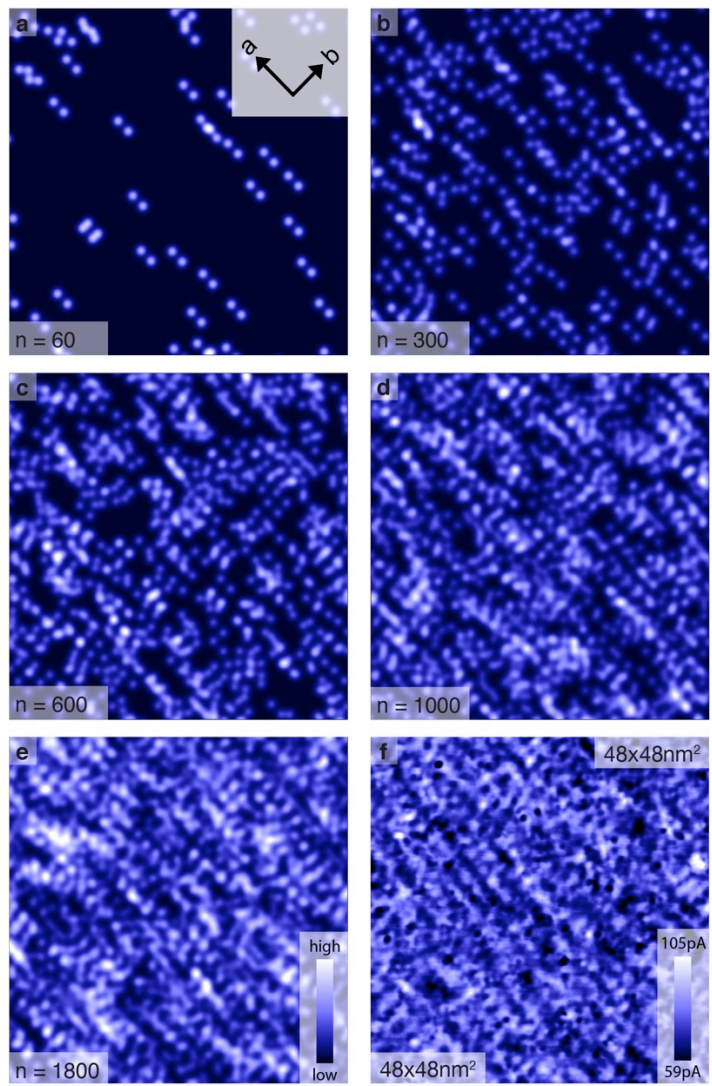

**Figure S3** Comparison between simulations of randomly scattered anisotropic impurity states and our measurement. The images are similar to Fig. 2*F*, but with different numbers of anisotropic impurity states: n=60, 300, 600, 1000 and 1800, which are shown from (a) to (e). (f) The measured the data as depicted in Fig. 2*B*.

## V. Identification of cobalt dopant atoms

In $Bi_2Sr_2CaCu_2O_{8+\delta}$, each oxygen dopant is negatively charged and generates an electronic impurity state below the Fermi level at ~-0.96eV 7. In the case of Co-doped $CaFe_2As_2$, each Co atom is substituting a Fe atom; it was predicted to create a dopant-induced impurity states at lower energies compared to the cuprates, and above the Fermi level 8. To search for such states, $g(r, E=eV)$ images are acquired from E=0 to E=+250meV. A conductance image, $g(r, 150meV)$, shown in Fig. S4B is taken in the same field of view as in Fig. S4A. It shows randomly distributed bright maxima that can be observed in $g(r,E)$ maps from about E=+100meV to +250meV; we identify them as Co sites. Some maxima in $g(r,E)$ at the sites that are also bright regions in the topograph (a few examples are circled in Fig. S4A), come from Ca clusters or other positively charged molecules on the surface, and are excluded in our dopant identification. Fig. S4C shows the location of Co atoms in the same FOV. Each Co atom is represented by a black circle with Gaussian function. However, this identification exhibits only ~20% of the density expected in the samples as compared to the average measured Co density by Wavelength dispersive spectroscopy, for reasons not yet resolved. As expected, these bright objects and nanoscale electronic structures are not observed in the pure $CaFe_2As_2$.

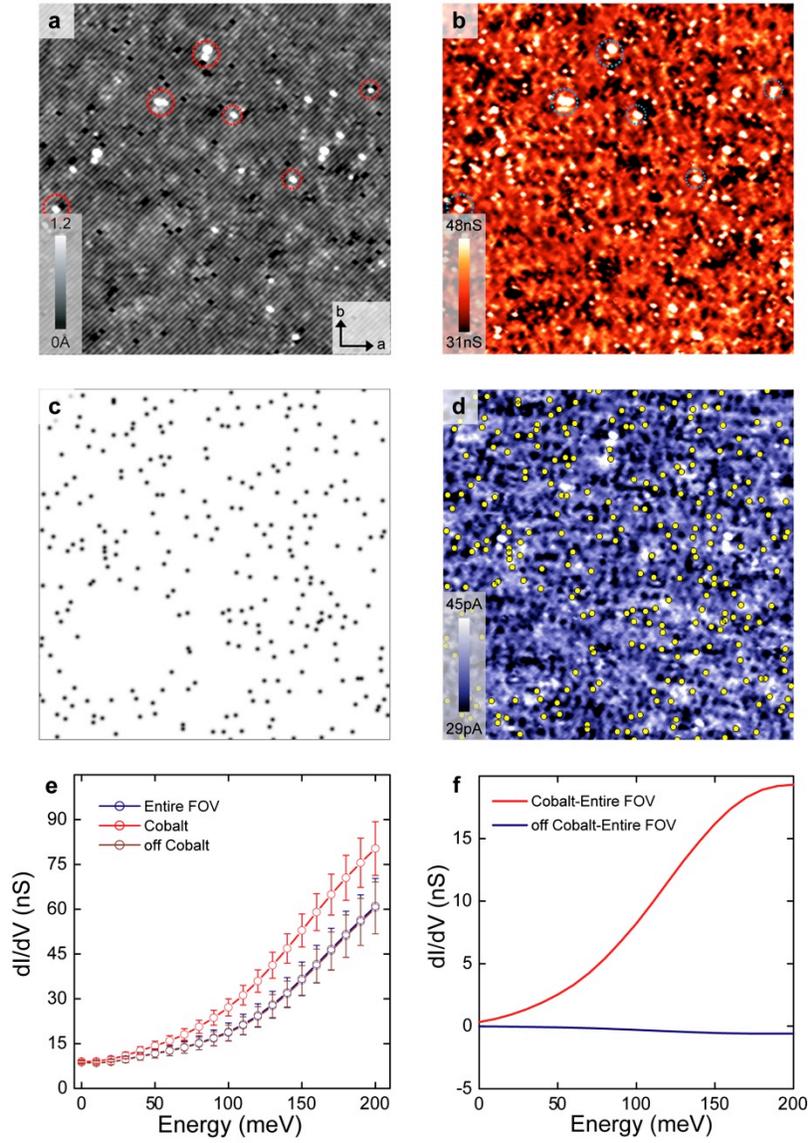

**Fig. S4** (*A*) Topography of Ca(Fe$_{1-x}$Co$_x$)$_2$As$_2$ taken in a 61.3×61.3nm field of view at $V_{Bias}$=+50mV and $I_{Setpoint}$=10pA. (b) A simultaneous conductance map, $g(r, E = 150 meV)$, is taken in the same FOV as Fig. S4*A*. (c) An artificial image shows the location of Co atoms (black circles) after the identification. (d) The Co atom positions (yellow circles) are overlaid with $I(r, E = 50 meV)$ image to show a possible correlation between Co and the nematic electronic structure, c.f. Fig. S5. (e) The average spectrum taken from the entire FOV, the Co positions and the positions away from Co atoms. (f) The average spectrum on the Co positions and the positions away from Co atoms after subtracting the average spectrum of whole FOV, yielding enhanced spectral weight at the Co position as predicted by Ref. 8.

## VI. Relating cobalt sites and anisotropic impurity states

Testing the local correlation between anisotropic impurity states and Co positions is challenging since there exist a large number of anisotropic impurity states that are apparently overlapping, far more than visible Co signatures. As a consequence, the surrounding of every visible Co atom might show the combined signal of several anisotropic impurity states, as well as other spatial inhomogeneities. Thus, even if there were an anisotropic impurity state centered on each Co site, one would not see it. Consequently, we chose to use a novel method to address this problem: we averaged the local surrounding of each Co position to remove noise, differences in local environments, and overlapping anisotropic impurity states, and to retain only the persistent characteristics of the local electronic structure characteristic for a Co atom.

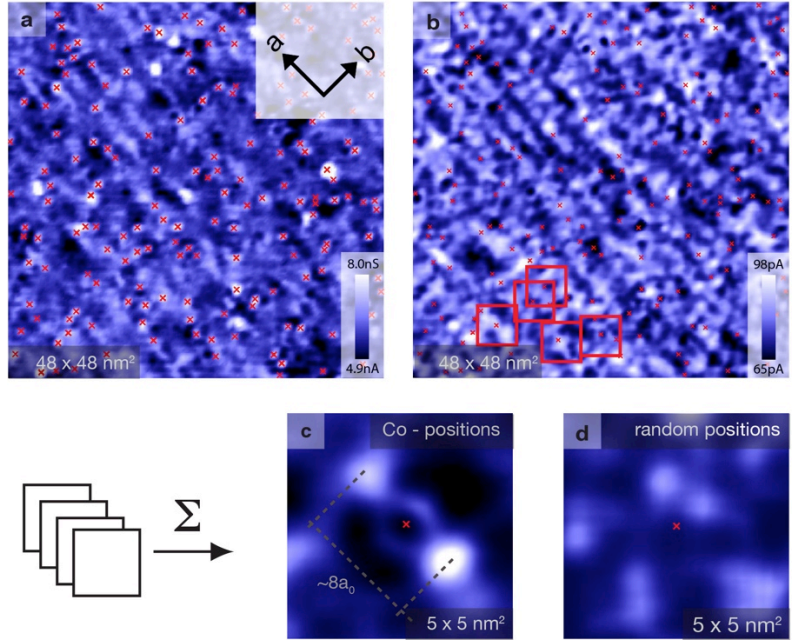

**Fig. S5** Testing the local correlation between anisotropic impurity states and Co positions. (a)-(b) Conductance and current images taken at 110 and -37 meV to image the Co-positions and anisotropic impurity states, respectively. The exact positions of the Co atoms are obtained by two-dimensional Gaussian fits on $g(r, E = 110 meV)$8 and are marked with red crosses. (b) On the current image that exhibits the anisotropic impurity states, $I(r, E = -37 meV)$, a small, 5nm square FOV is cropped out around every Co position (some are marked by red boxes). The average of all Co positions, $\langle I(r, E = -37 meV) \rangle_{Co}$ is shown in (c). (d), The same procedure applied on random positions instead of Co positions yields random noise.

Fig. S5 illustrates the process. The positions of Co-signatures on the conductance map at 110meV[8] are determined by local two-dimensional Gaussian fitting (c.f. Section V), and transferred to the atomically registered current image at -37meV that exhibits the anisotropic impurity states. We then crop out a small ~5nm square FOV around each position and average over all 230 such squares to obtain the persistent local electronic environment. This is shown in Fig. S5*C*. Indeed, an anisotropic impurity states is found centered at the Co position. As a test, we also performed the same procedure as described above, but at random positions instead of at Co positions. This yields random noise, shown in Fig. S5*D*.

## VII. Comparison of different *q*-vectors observed in QPI pattern

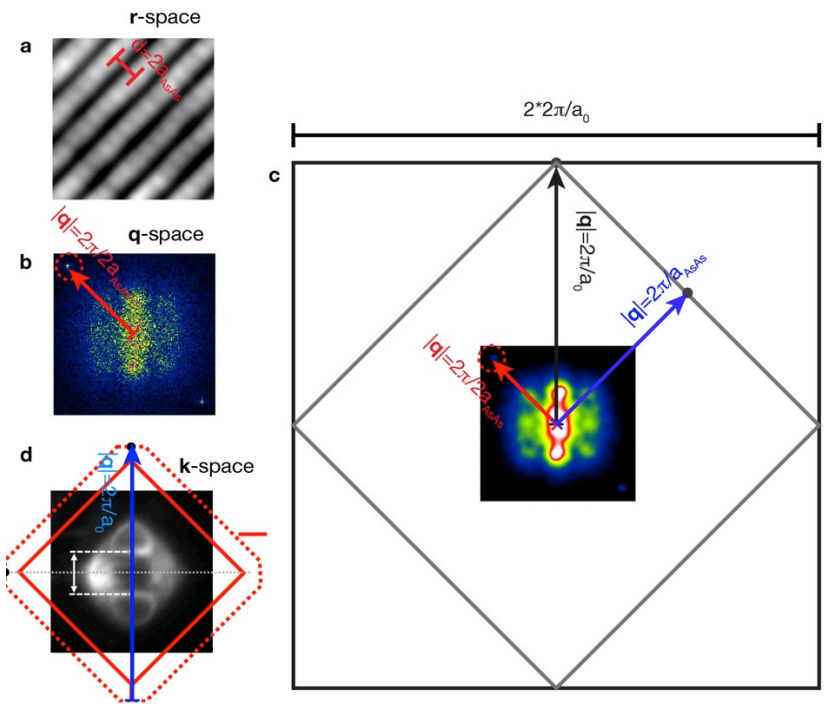

**Fig. S7** Comparison of different *q*-vectors in QPI patterns. (a) The surface reconstruction in the topography gives us a length scale of $2a_{AsAs}$. (b) The Fourier transform of conductance images (here $g(q, E = 0meV)$) show a peak corresponding to this. (c) Overview of all relevant *q*-vectors in the problem. As in the main text, $a_0$ is the shortest Fe-Fe distance. (d) Comparison with *k*-space (reproduced from Ref. 2).